\documentclass[a4paper,11pt]{article}
\pdfoutput=1 

\usepackage{jinstpub} 

\title{\boldmath Neutrino detectors for oscillation experiments}

\author[a,b,c]{Yury Kudenko}

\affiliation[a]{Institute for Nuclear Research of RAS,\\60 October Revolution Pr 7a,
Moscow, Russia}
\affiliation[b]{Moscow Institute of Physics and Technology, 
\\Institutskiy per. 9, Dolgoprudny, Moscow Region, Russia}
\affiliation[c]{Moscow  Engineering Physics Institute,\\Kashirskoe shosse 31, Moscow,  Russia}

\emailAdd{kudenko@inr.ru}

\abstract{A brief overview  of the  development of neutrino detectors for long-baseline oscillation experiments at accelerators and reactors is presented. Basic principles and main features of detectors  of running  accelerator experiments T2K and NO$\nu$A sensitive to  a first level of  CP violation and neutrino mass hierarchy, and reactor experiments Daya Bay, RENO and Double Chooz which measured the mixing angle $\theta_{13}$ are discussed. A variety of different experimental techniques is proposed and developed for the next generation oscillation experiments: a 20 kt scintillator detector for the reactor experiment JUNO, a 0.52 kt water-Cherenkov detector Hyper-Kamiokande, and  a massive liquid argon time-projection chamber neutrino detector envisaged for the DUNE experiment. Present status of these detectors, recent progress in R\&D and future prospects are summarized in this paper. }

\keywords{Neutrino detectors, Cherenkov detectors, liquid scintillators, noble liquid detectors, time projection chambers, photomultipliers.}

\collaboration[c]{}

\proceeding{International Conference "Instrumentation for Colliding Beam Physics"  (INSTR17) \\
  27 February - 3 March, 2017 \\
  Novosibirsk, Russia}

\begin{document}
\maketitle
\flushbottom

\section{Introduction}
\label{sec:intro}
The discovery of neutrino oscillations has provided convincing evidence for non-zero neutrino masses and leptonic mixing. This phenomenon  is the first clear example of new physics beyond the Standard Model~(SM). Oscillation probabilities between the three active neutrinos are described by six independent parameters:  3 mixing angles $\theta_{12}$, $\theta_{23}$, and $\theta_{13}$,    two neutrino mass-squared 
differences  $\Delta m^2_{21} = m^2_2 - m^2_1$, $\Delta m^2_{32} = m^2_3 - m^2_2$, and possibly  one complex CP violating phase $\delta_{CP}$.   These parameters  are determined in oscillation  experiments with atmospheric, solar, reactor and accelerator neutrinos, except for $\delta_{CP}$ and the sign of $m^2_{32}$.    Both the normal mass hierarchy ($m_3 \gg m_2 > m_1 $) and inverted hierarchy ($m_2 > m_1 \gg m_3 $) are possible. The science program of current and future long baseline  accelerator and reactor experiments is focused on a search of CP violation in leptonic sector and measurement of $\delta_{CP}$, determination of neutrino mass hierarchy, and precision measurements of  mixing parameters.  The concepts of these experiments as well as  various   detector techniques will be described below.  


\section{Current experiments}
\subsection{T2K}
The T2K (Tokai-to-Kamioka) experiment~\cite{Abe:2011ks} uses a high intensity off-axis neutrino beam produced by 30 GeV protons at J-PARC (Japan Proton Accelarator Research Complex). The far detector,  Super-Kamiokande, a well known 50 kt Cherenkov detector located at a distance of 295 km from J-PARC, measures the spectra of muon and electron neutrinos (antineutrinos) after oscillations. Super-Kamiokande  distinguishes muon and electron  from neutrino charged current interactions  using the shape of the  Cherenkov ring. The reconstruction of the neutrino energy  is made with lepton kinematics assuming a charged-current quasi-elastic scattering, in which a neutrino is converted to a charged lepton in the reaction $\nu_l +n \to l^- + p$. The T2K  $2.5^{\circ}$ off-axis beam has a relatively narrow band   neutrino (antineutrino)  flux with a peak energy of 0.6 GeV,  which   corresponds to the first oscillation maximum at Super-Kamiokande. The T2K near detector complex located at 280 m from the pion production  target  consists of an on-axis detector (INGRID) and a  set of multiple sub-detectors (ND280)  placed at the  angle of $2.5^{\circ}$ relative to the proton beam direction inside the UA1 magnet operated with a magnetic field of 0.2 T. The near detectors designed to characterize the     parameters of the unoscillated neutrino beam and neutrino cross sections. The tracker   consists of two fine-grained detectors (FGDs) and three time projection chambers (TPCs) and provides the primary input data in the current T2K oscillation analyses. Both FGDs (the first detector is comprised of scintillator bars, the second detector consists of scintillator bars and water layers) are utilized as a neutrino target. Three TPCs are used to measure the momenta and the sign of charged particles. The neutrino flux and cross-section models  are constrained using data collected in  ND280.
 
\begin{figure}[ht]
\centering 
\includegraphics[width=14cm]{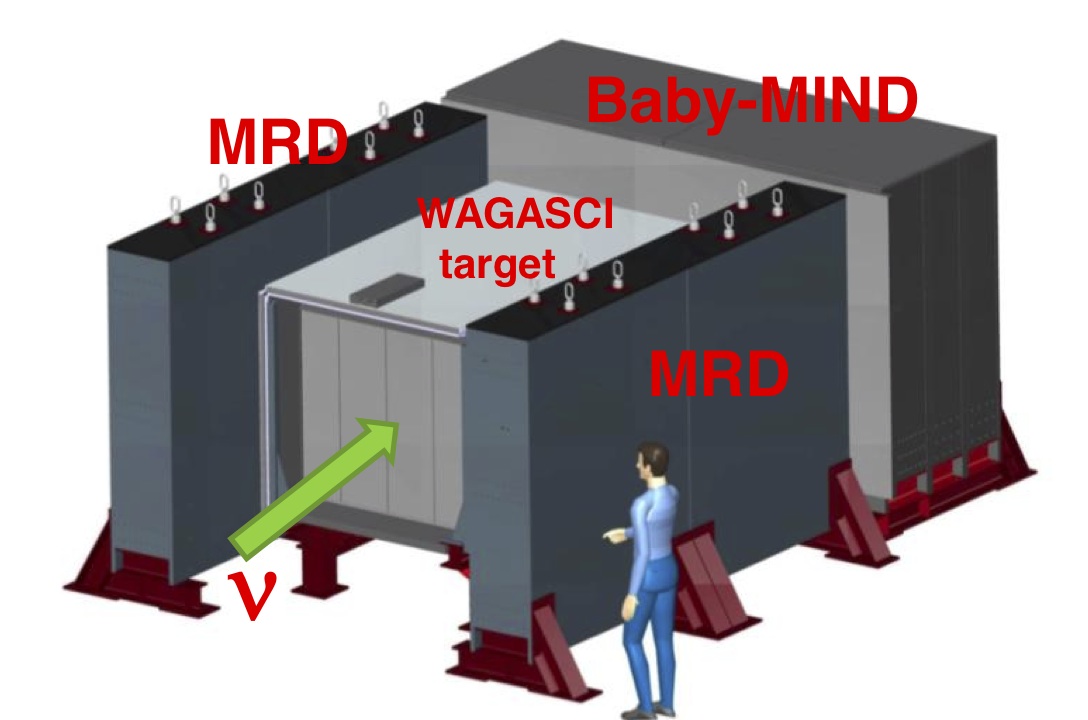} 
\caption{\label{fig:wagasci-babymind} The layout of the WAGASCI set-up.}
\end{figure}

\begin{figure}[hb]
\centering 
\includegraphics[width=.6\textwidth]{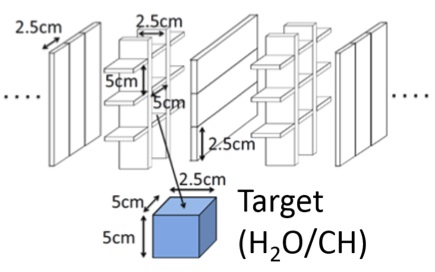}
\includegraphics[width=.35\textwidth]{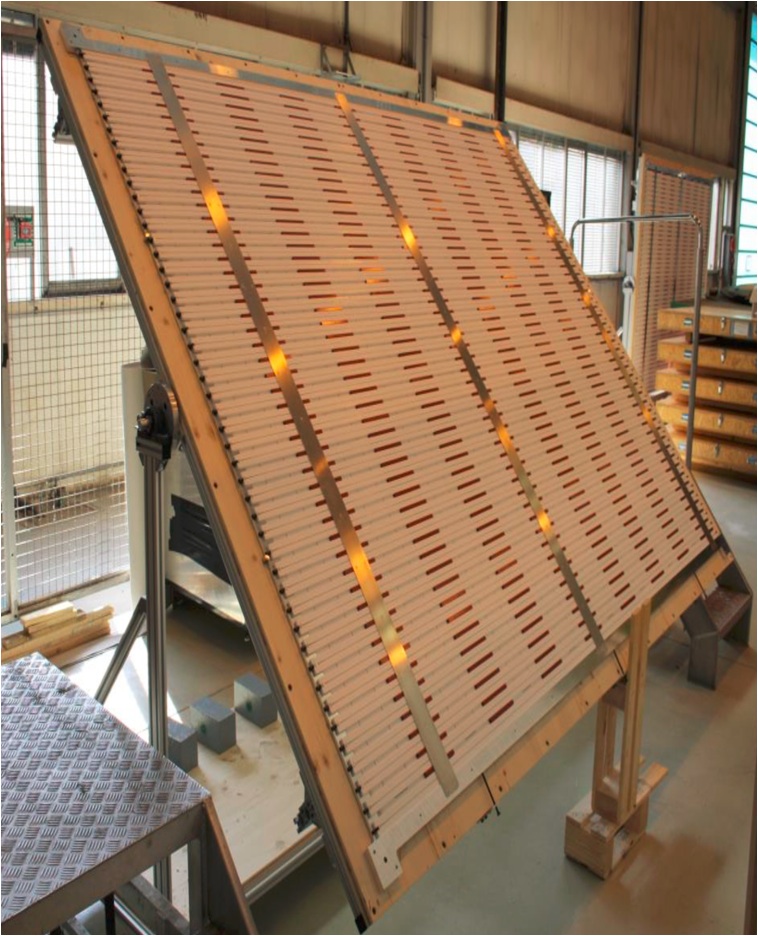}
\caption{\label{fig:target-plane} Left panel: A novel 3D grid-like structure of the plastic scintillator of the neutrino target. Cells are filled with either H$_2$O or CH. Right panel: A scintillator plane of the  Baby-MIND. }
\end{figure}

The T2K Collaboration has successfully applied a method of fitting the near detector data with parametrized models of the neutrino flux and interaction cross sections and reduced the systematic uncertainties on the predicted rate of $\nu_{\mu}$ and $\nu_e$ at SK to a 6-7\% level~\cite{Abe:2015awa}.  However, there are several limitations to this approach. In the T2K experiment, water is used as the neutrino target in SK. But the mass fraction of water of the second FGD is only 47\%. The ND280 detector has a low efficiency for large angle tracks while Super-Kamiokande has a 4$\pi$ angular acceptance.  The uncertainty of the neutrino cross section on water due to this difference is one of the major systematic errors in the T2K neutrino oscillation analysis.   To measure neutrino (and antineutrino) cross sections on water and hydrocarbon with high precision and large angular acceptance a new neutrino detector, WAGASCI (WAter Grid And SCIntillator)   has been developed~\cite{Koga:2015iqa}. The goal of this experiment is to  measure the ratio of neutrino cross sections on plastic and water with a  3\% accuracy at the neutrino energies $\leq 1$ GeV. The WAGASCI experiment will utilize the detector which comprises a  segmented  target of water and scintillator cells, muon range detectors, and a magnetic spectrometer  to measure momentum and charge identification of the outgoing muons from charged current interactions.  Fig.~\ref{fig:wagasci-babymind}  
shows the  layout of the WAGASCI set-up, with the active neutrino  target WAGASCI in  the center, the Magnetized Iron Neutrino Detector(Baby-MIND)~\cite{Antonova:2017thk}, and  two side Muon Range Detectors (MRDs). The central neutrino interaction target consists of 4 blocks, each $1\times 1\times 0.5$ m$^3$, with an alternating pattern along the beam direction between water-filled and hydrocarbon-filled blocks instrumented with  plastic scintillator detectors, as shown in Fig.~\ref{fig:target-plane}.
These detectors will be read out with arrays of 32 channel Multi Pixel Photon Counters (MPPCs). MRDs which  measure the momentum of muons by range consist of several plates of steel each of  3-cm thick, interleaved with plastic scintillator detector modules. The Baby-MIND detector consists of 33 magnetized 3-cm steel plates which provide a magnetic field of 1.5 T and 18 scintillator modules that measure the position of hits along the spectrometer and the curvature of the track in the magnetic field. Detector modules consist of x-y planes of plastic scintillator bars (Fig.~\ref{fig:target-plane}), with wavelength shifting fibers read out with single MPPCs.  The  WAGASCI detector is expected to be commissioned and start  data taking in the beginning of 2018.  

\subsection{NO$\nu$A}
\begin{figure}[htbp]
\centering 
\includegraphics[width=14cm]{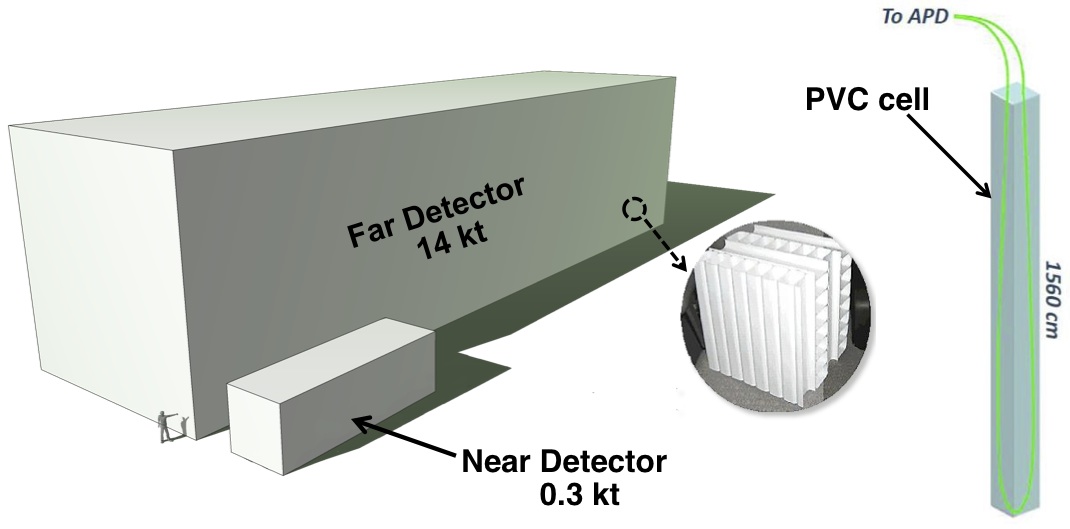} 
\caption{\label{fig:nova_detector} Schematic view of the far and near NO$\nu$A detectors. The inset shows the orthogonal orientation of adjacent layers. A PVC cell filled with the  liquid scintillator equipped with a WLS  fiber is also shown.}
\end{figure}

The NO$\nu$A (NuMI Off-axis $\nu_e$ Appearance) experiment~\cite{Adamson:2016tbq}  is designed to study neutrino oscillations using Fermilab's NuMI neutrino beam. The results obtained in this experiment are consistent with  the T2K $\nu_e$ appearance measurements and confirmed obtained in T2K  constraints on possible $\delta_{CP}$ with a preference for $\delta_{CP} \simeq -\pi/2$. NO$\nu$A has a long baseline of 810 km that leads to better sensitivity of mass hierarchy. NO$\nu$A uses two neutrino  detectors: a 300 t near detector  located approximately 1000 meters downstream of the pion production target and 100 m underground, and  a 14 kt far detector  is situated on the surface of the earth  in Ash River, Minnesota.  Both liquid scintillator-based detectors with identical structures (Fig.~\ref{fig:nova_detector}).
are located at the angle of  14.6 mrad from the central axis of the NuMI beam. The neutrino beam with  a narrow energy interval centered at 2 GeV  is tuned  to the first oscillation maximum with "atmospheric" oscillation parameters.  The NO$\nu$A detectors are  highly segmented  tracking calorimeters.  The segmentation  is provided by a lattice of  extruded  polyvinyl chloride (PVC) cells with a cross section of $6.6 \times 3.9$ cm$^2$. The cells are $\sim 15$ m long in the far detector and $\sim 4$ m long in the near detector. Each scintillator-filled  PVC cell is equipped with a WLS Y11 fiber looped at the far end of the cell. Both near ends of the WLS fiber terminate on the same pixel of a 32-pixel avalanche photodiode (APD). The far detector has about  $3.5\times 10^5$ PVC cells. Approximately 37\% of the detector mass is in the PVC structure and 63\% of the mass is in the liquid scintillator. The radiation length of  the detectors (38  cm) is much larger than the cell dimensions. The geometry of NO$\nu$A detectors allows to provide good reconstruction and separation of electromagnetic and hadronic   showers, identification of muons and neutral pions. The experiment is taking data with the beam power of 700 kW and an  exposure of $36\times 10^{20}$ protons on target is expected to be delivered by  2023. 

\subsection{Reactor experiments}
\label{subsec:reactor_experiments}
The main goal of three running reactor experiments Daya Bay~\cite{An:2012eh} in China, RENO~\cite{Ahn:2012nd} in Korea, and Double Chooz~\cite{Abe:2011fz} in France is the precise measurement of the third mixing angle  $\theta_{13}$. To accomplish the task,  these  experiments use a near-far detector conception in that a  near detector is located close to a reactor where the oscillation due to $\theta_{13}$ is small and a far detector is installed at  $\sim 1$ km from the reactor, an optimal distance  to provide a good sensitivity to  $\theta_{13}$. All neutrino detectors in these experiments use a liquid scintillator doped with gadolinium. Since both  detectors have identical structure, the systematic uncertainties  of neutrino flux, neutrino cross section, detection  efficiency, and detector mass  are significantly suppressed. The experiments use the inverse beta decay  to detect reactor antineutrinos   in a liquid scintillator  
\begin{equation} 
\bar{\nu}_e + p \to e^+ + n. 
\label{eq:reactor_nu_detection}
\end{equation}
The cross section of this process is very well known.  The positron deposits energy in the scintillator, which is converted  in   scintillation light. The positron then  annihilates with an electron with a contribution 1.02 MeV to its kinetic energy.  The energy of the detected reactor antineutrino can be obtained from the energy of the positron signal $E_{\bar\nu_e} = E_{e^+} + 0.8 ~{\rm MeV}$. The neutron quickly thermalizes  and is absorbed by Gd within tens  $\mu {\rm s}$. Then the exited Gd isotope emits photons with the total energy $\sim 8$ MeV. So the antineutrino signal is identified by the delayed coincidence of the positron signal and the Gd signal. RENO and Double Chooz use two detectors, one near and one far. The Daya Bay experiment has a more complex set-up: six nuclear reactors with a total power of 17 GW$_{\rm th}$ (thermal power), four near detectors located in two places close to reactors, and four far detectors, each  contains 20 t of liquid scintillator. The energy resolution of 8\%/$\sqrt{E(\rm MeV)}$ was obtained  in Daya Bay detectors.   The main oscillation result, the value of $\theta_{13}$ was obtained from  the ratio of antineutrinos detected in the far and the near detectors. It is independent of the calculation of the antineutrino flux and the spectrum from the reactors.  After a few years of data taking, Daya Bay have measured ${\rm sin}^22\theta_{13}$ with a  very impressive precision of about 2.5\%.  

\section{Future projects}
\subsection{JUNO}
\begin{figure}[htbp]
\centering 
\includegraphics[width=15cm]{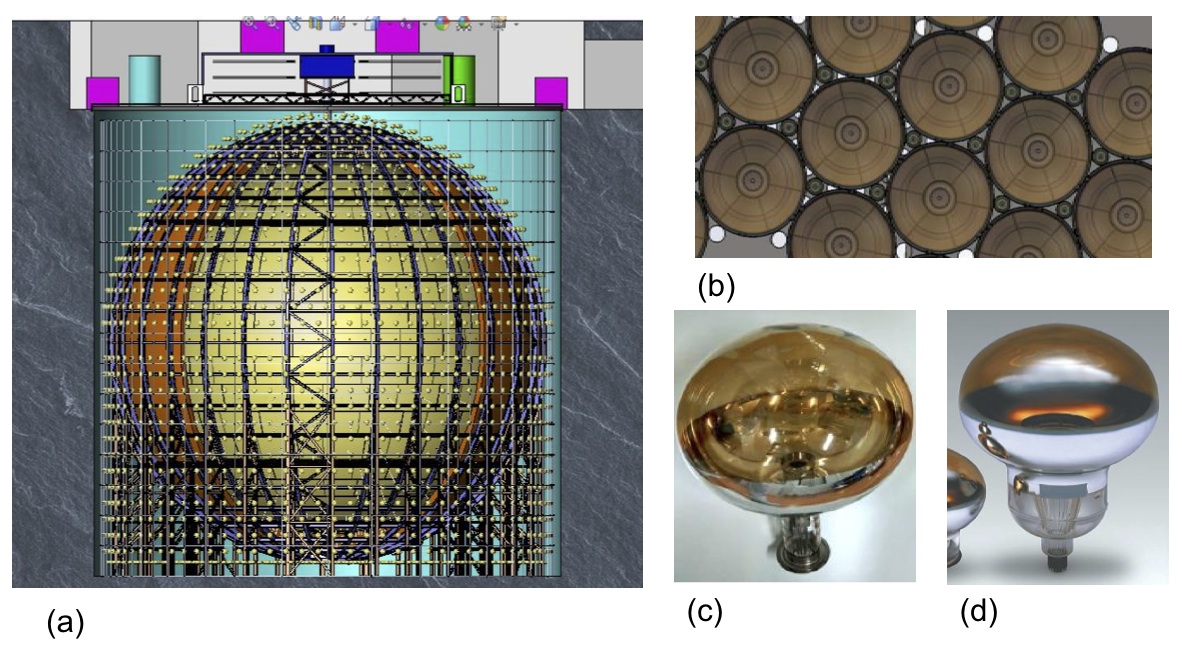} 
\caption{\label{fig:juno}Schematic view of the JUNO central detector (a), the PMT grid  comprised of large and small PMTs~(b), a 20 inch MCP-PMT~(c), and a 20 inch Hamamatsu PMT~(d).}
\end{figure}

The Jiangmen Underground Neutrino Observatory (JUNO)~\cite{An:2015jdp},  a large liquid scintillator detector located at a 53 km distance from reactors, has a  primary goal of the determination of the neutrino mass hierarchy. The detector will be located 1800 meters of water equivalent~(m.w.e.) underground and consists of a 20 kiloton liquid scintillator (based on LAB doped with PPO and bis-MSB) contained in a huge 35.4 m diameter acrylic sphere (Fig.~\ref{fig:juno}(a)), 
instrumented by more than 17000 20-inch  and about 34000 3-inch PMTs. Small PMTs will be  mounted in gaps between  large phototubes, as shown in Fig.~\ref{fig:juno}(b) ensuring a 77\% photocathode coverage. It should be noted that the scintillator does not contain Gd  in order to be more stable in time. To discriminate between the neutrino hierarchies at $\geq 3\sigma$, the  energy resolution is required to be  3\%/$\sqrt{E(\rm MeV)}$ and the absolute energy scale should be calibrated with a precision of 1\%. To meet these criteria  and provide a high light yield of $\sim 1200$ p.e./MeV, PMT's should have a quantum efficiency $\geq 27$\% and the attenuation length of the liquid scintillator should be longer than 20 m at 430 nm~\cite{heng_instr17}.  The antineutrino event will be identified observing the inverse beta decay~(\ref{eq:reactor_nu_detection}) and the delayed signal of 2.2 MeV from the neutron captured by hydrogen within 200 $\mu$s. Two types of 20" PMTs will be utilized: superalkali Hamamatsu PMTs (Fig.~\ref{fig:juno}(d)) and new MCP-PMTs (Fig.~\ref{fig:juno}(c)) in which the dynode chain is replaced by micro channel plates. A new design of MCP-PMTs integrates two photocathodes, a transmission and reflective ones~\cite{Sen:2016kqy}. This approach allows to recover some nonconverted photons. Both PMT types reached the quantum efficiency more than 30\%  that is needed to obtain the required energy resolution. JUNO  plans to start data taking in 2020.

\subsection{Hyper-Kamiokande}
\begin{figure}[htbp]
\centering 
\includegraphics[width=14cm]{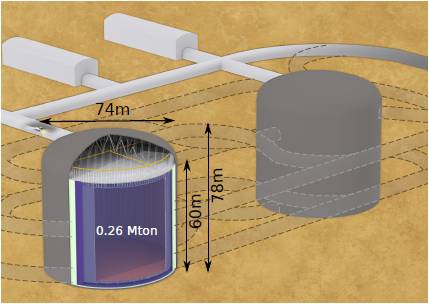} 
\caption{\label{fig:hyperk} Schematic view of the Hyper-Kamiokande.}
\end{figure}

\begin{figure}[htbp]
\centering 
\includegraphics[width=14cm]{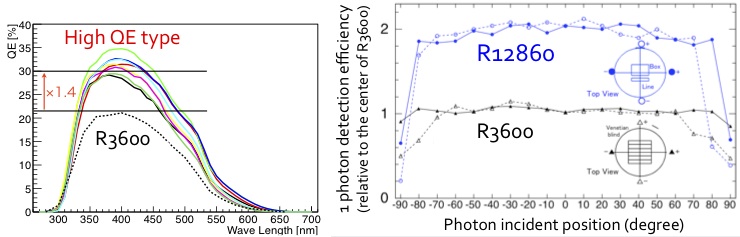} 
\caption{\label{fig:hyperk_pmt}The quantum efficiency of Hamamatsu R12860 (left) and the single photon detection efficiency as  a function of incident positions (right) of the new Hamamatsu PMT R12860  developed  for Hyper-Kamiokande~\cite{Nakayama:2016vdx}. Also shown parameters of  the Super-Kamiokande  Hamamatsu  PMT R3600.  }
\end{figure}

The Hyper-Kamiokande detector (HK)~\cite{Abe:2015zbg}, a proposed next-generation water-Cherenkov detector, will have a broad physics program which covers many areas of particle and astroparticle physics. It will serve as a far detector of a long-baseline neutrino experiment T2HK (Tokai-to-Hyper-Kamiokande)  using  neutrino and antineutrino  beams produced at  J-PARC upgraded to a $\sim 1$ MW of power. The main goal of the T2HK experiment is the sensitive search for CP violation in neutrino oscillations~\cite{Abe:2014oxa}, i.e. the observation of  the CP violation with a significance of $\geq 5\sigma$ for about 60\% of all values of $\delta_{CP}$ and the measurement of $\delta_{CP}$ with a precision of $7^{\circ}$ for  $\delta_{CP} = 0^{\circ}$ or about $20^{\circ}$ for  $\delta_{CP} = -\pi/2$ in 10 years of data taking. To achieve this goal the total systematic uncertainty on  far detector rate prediction should  be less than 3\%. This issue can be addressed by the upgrade of ND280 and by building an intermediate water Cherenkov detector (see, for example \cite{Bhadra:2014oma})  at about 1 km from the T2K neutrino beam production target.  Hyper-Kamiokande  will also increase existing sensitivity to proton decay by an order of magnitude, and  will study neutrinos from various sources, including atmospheric neutrinos, solar neutrinos, supernova neutrinos and annihilating dark matter. HK is based on the proven technology of the  highly successfull Super-Kamiokande detector.  The HK detector (Fig.~\ref{fig:hyperk}) 
will be located about 8 km south of Super-Kamiokande  with  an overburden of 1750 m.w.e.   It will consist of two cylindrical tanks (each 60 m high and 74 m in diameter) and have a total (fiducial) mass of 0.52 (0.37) Mton, making it 10 (17) times larger than its predecessor Super-Kamiokande. It will use 40,000 PMTs per tank to reach the same 40\% photocoverage.  Similar to Super-Kamiokande, an outer detector with the layer width  of 1-2 m will help to constrain the external background.  In order to achieve  broad scientific goals, particles with a wide range of energies should be  reconstructed. The number of Cherenkov  photons that hit each photosensor ranges from one to several hundred. Thus, the photosensors are required to have a high photon detection efficiency, a  wide dynamic range, and a good linearity. The location of the neutrino interaction vertex is reconstructed using the Cherenkov photon arrival timing information at each PMT. Therefore, good timing resolution of  photosensors is essential, and the jitter of the transit time is required to be less than 3 ns ($1\sigma$) for a single photon. To meet these requirements,  two types of new 50 cm diameter vacuum-based photodetectors have been developed for Hyper-Kamiokande. One is the  Hamamatsu PMT R12860   with a box-and-line dynode, which has a faster time response and a better collection efficiency compared to  Hamamatsu PMT  R3600 that have been used successfully in Super-Kamiokande. The other is a hybrid photodetector, HPD, which uses an avalanche diode  as an electron multiplier. Hamamatsu R12860 has a very good  timing and charge resolutions for the single photon,  $\sim 1$ ns  and about 35\%, respectively.  An improved photocathode of R12860 allows to reach  the quantum efficiency of 30\% at 400 nm, about 1.4 times higher than that of the Super-Kamiokande  PMTs, as shown in Fig.~\ref{fig:hyperk_pmt}~(left), 
and  the photoelectron collection efficiency of R12860 is also much higher.   As a result,  the total efficiency for the single photon detection of Hamamatsu R12860 is almost twice higher than that of the Super-Kamiokande PMTs, as shown in Fig.~\ref{fig:hyperk_pmt}~(right). This newly developed high-efficiency and high-resolution  PMTs met the requirements for the HK photosensor~\cite{berardi_instr17} and will significantly contribute to the main goal of the Hyper-Kamiokande experiment. It should be also mentioned that R12860 has an improved pressure tolerance so that a deeper tank becomes feasible. Alternative solutions for photosensors are also extensively studied by the international collaboration~\cite{berardi_instr17}. In case of approval, the  construction of Hyper-Kamiokande  can start in 2018 and data taking is expected to begin  in 2026.

\subsection{DUNE}
\begin{figure}[htbp]
\centering 
\includegraphics[width=14cm]{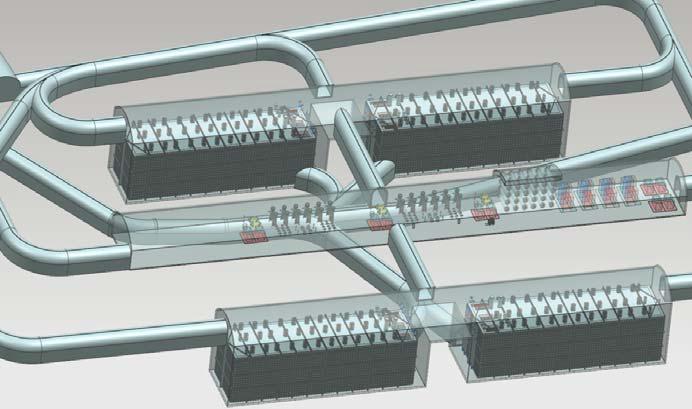} 
\caption{\label{fig:dune_4detectors} DUNE far detector complex: 4 liquid Ar TPC modules with approximate dimensions $15(\rm width) \times 14(\rm height) \times 60(\rm length)$ m$^3$. Each module is able to contain a total mass of about 17000 t of liquid argon.}
\end{figure}

The main scientific goals of the Deep Underground Neutrino Experiment~(DUNE)~\cite{Acciarri:2015uup} are  the  sensitive test of  CP violation in the leptonic sector, determination the neutrino mass hierarchy, and precise measurements of neutrino oscillation parameters. The proposed far neutrino detector  will be built deep underground, at a depth of about 1500 m,  in the Sanford Underground Research Facility (South Dakota, USA), about 1300 km from Fermilab where a high intensity wide band on-axis neutrino beam with neutrino energies of 1-6 GeV will be formed. The near detector baseline design is a fine-grained tracker  in a magnetic field, complemented by calorimetry and muon range detector. It will be located at a distance of 574 m from the target, at a depth of 65 m. The far detector will  consist of four cryostats instrumented with Liquid Argon Time Projection Chambers (LAr TPCs) with a fiducial mass of 40 kt~\cite{Acciarri:2016ooe}. The LAr TPC technology offers excellent capabilities for position and energy resolution and for high-precision reconstruction of complex interaction topologies over a broad neutrino  energy range and  will provide a powerful complementarity to the large, underground water Cherenkov or scintillator-based detectors. It is expected that DUNE will reach a $3\sigma$ CP coverage of 75\% and a $5\sigma$ coverage of 50\% for an exposure of about 1 Mt$\times$MW$\times$year (20 years of running  with the 40 kt detector and a 1.2 MW proton beam power).  DUNE will determine neutrino mass hierarchy with a sensitivity $\geq 5\sigma$ for 100\% of $\delta_{\rm CP}$ values for 7 years of data taking (3.5 years in neutrino mode + 3.5 years in antineutrino mode). These sensitivities still depend on the allowed range of values of other neutrino oscillation parameters.   It is assumed that all four detector modules (Fig.~\ref{fig:dune_4detectors}) will be similar but not necessarily identical. 
According to the  present timescale, DUNE begins  data taking with the first 10 kton module in 2026 and the full configuration will be ready by 2029. 

\begin{figure}[hb]
\centering 
\includegraphics[width=15cm]{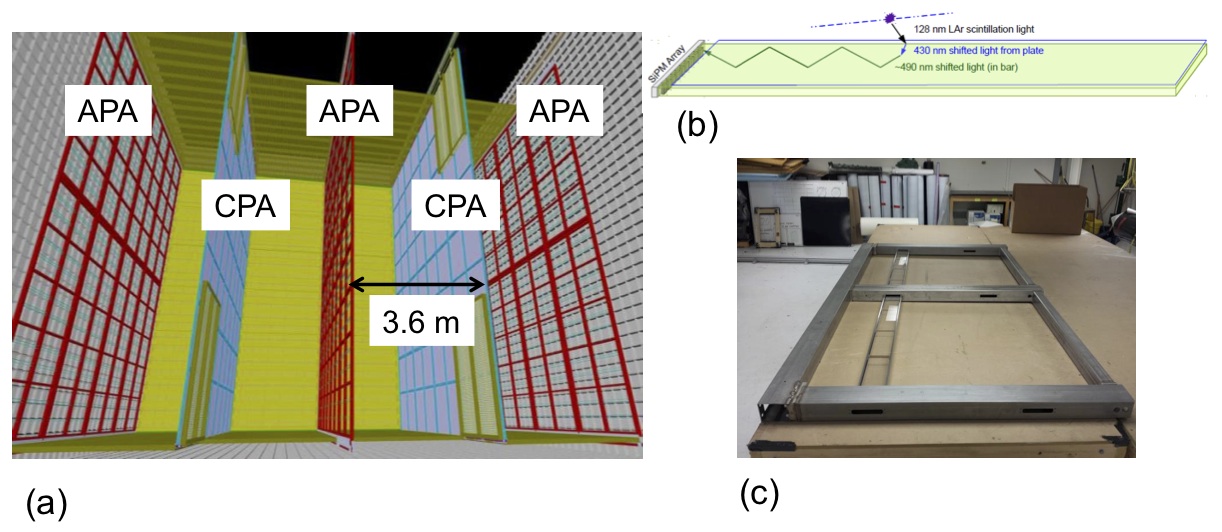} 
\caption{\label{fig:dune_sp_photodet}A view of the partially installed single-phase TPC inside the membrane cryostat (a), a single photon detector module (b), and a PD module mounted on the APA frame (c)~\cite{manenti}.}
\end{figure}

The first DUNE liquid argon module  adopts a single-phase technology pioneered by ICARUS T600  detector. The single-phase TPC will be  constructed by placing alternating high-voltage  cathode planes and anode readout planes with a 3.6 m spacing in a bath of ultra-pure liquid argon. In this design, the charge is collected directly without gain, enabling precision charge calibration, but  signal levels are low that motivates locating the front-end electronics in the LAr close to anode wires. This innovating approach  requires the use of cold electronics.  According to the reference design, the novel photon detection (PD) system will be comprised of inserted into anode plane assembly~(APA) frame  wavelength shifting bars with SiPM readout.   A prototype of the single-phase TPC (700 t LAr TPC) will be constructed at CERN using 6 full size APAs with fully instrumented electronics and 6 cathode plane assemblies~(CPAs).  Its geometry is $6~{\rm m} \times 2.3~{\rm m}$ cathode and anode planes with 3.6 m spacing as shown in Fig.~\ref{fig:dune_sp_photodet}(a).
A reference  PD module consists of a light guide and 12 silicon photo-multipliers~(SiPMs), as shown in Fig.~\ref{fig:dune_sp_photodet}(b). The 128-nm photons from LAr interact with the wavelength shifter on the surface of the bar and the wavelength-shifted light, with a peak intensity around 430 nm, is re-emitted inside the bar and transported through the light-guide to SiPMs mounted at one end of the bar. The PD  modules are mounted on the APA frames (Fig.~\ref{fig:dune_sp_photodet}(c)). A few  alternative photon detector designs which demonstrated the ability to detect the LAr light are currently being considered for the PD system. Data from the beam test of the single-phase TPC with PD system  will  provide input into the final technology decision.

\begin{figure}[htbp]
\centering 
\includegraphics[width=14cm]{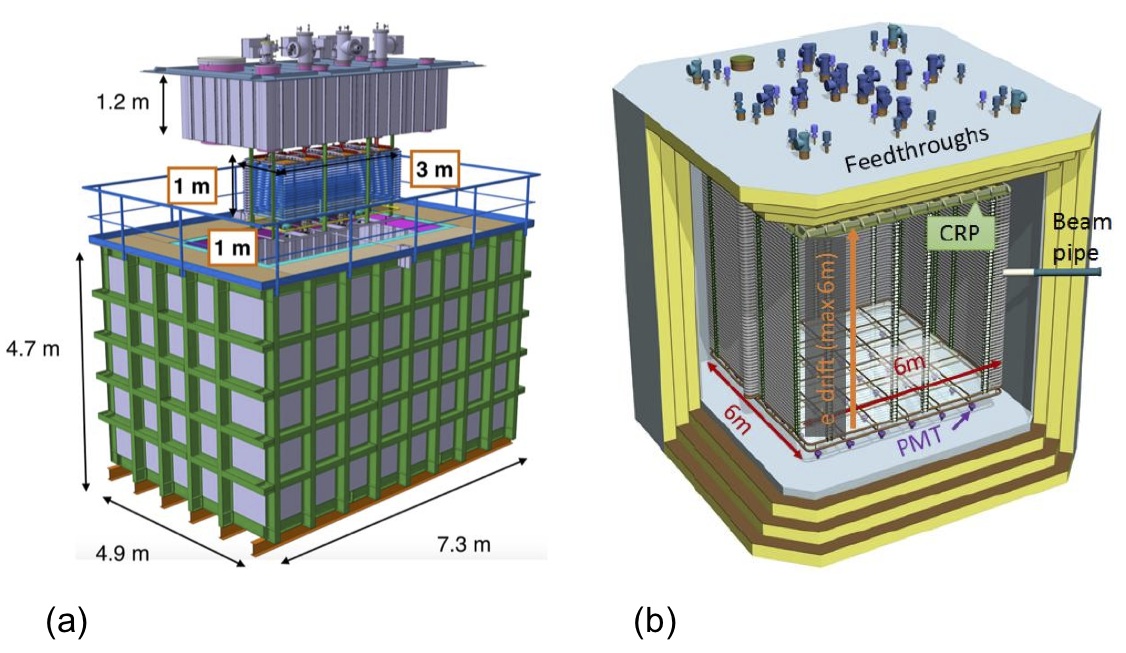} 
\caption{\label{fig:protodune}The 20-t dual-phase prototype~(a) and the $ 6\times 6\times 6$ m$^3$ WA105 dual-phase demonstrator~(b). }
\end{figure}

In the  dual-phase LAr TPC,   the ionization electrons are vertically drifted (drift distance is about 12 m)  in a constant electric field and extracted to a gaseous volume.   Once extracted, electrons  pass in the gas through a high electric field in the holes of a 1 mm thick LEM (Large Electron Amplifier) where they are amplified following avalanche cascades. The charge  is collected on a two-dimensional and  segmented anode. The expected signal-to-noise ratio is about $ 100:1$.  The prompt scintillation light is detected with an array of cryogenic PMTs and used to provide the trigger signal and the absolute time reference for the charge readout electronics.  The on-going LBNO-DEMO~(WA105) experiment at the CERN Neutrino Platform~\cite{cern_neutrino_platform} has to test all the principal components  of the dual-phase technology necessary for building a multi-kt scale detector. In this experiment, two LAr dual-phase LEM TPC detectors, shown in Fig.~\ref{fig:protodune},
will be built and tested~\cite{trzaska_inst17}. The performance of a  $ 3\times 1\times 1$ m$^3$ pilot TPC detector with the total LAr mass of 24 t (Fig.~\ref{fig:protodune}(a)) will be measured with cosmics. A $ 6\times 6\times 6$ m$^3$ (total LAr mass 700 t) demonstrator (Fig.~\ref{fig:protodune}(b)) will be exposed to particle beams at CERN. The second DUNE far detector module  can be constructed on the basis of  a dual-phase TPC if its performance is confirmed as a result of activities at the CERN Neutrino Platform.

The CERN Neutrino Platform provides the unique R\&D  and test facilities for DUNE  detectors and components. The performance of a full scale  single-phase  TPC module will be characterized and  a new technique for light detection in LAr will be learned. The functionality of cold TPC electronics will be verified. The technique of the dual-phase LAr TPC is expected to be proven.    The goal of beam tests at CERN of  the  single- and dual-phase detector prototypes includes measurements of detector response to hadronic and electromagnetic showers, study of ${\rm e}/\gamma$ separation, test  of event reconstruction algorithms etc.  The detectors are expected to start taking beam data at CERN in 2018.

\section{Summary}
Neutrino oscillations reveal physics beyond the Standard Model. Among many open neutrino questions, the discovery and study of leptonic CP violation, the determination of the neutrino mass hierarchy, and precision measurements  of the oscillation parameters require novel massive neutrino detectors. Current long baseline accelerator experiments T2K and NO$\nu$A can obtain a $3\sigma$ sensitivity to CP violation and to the  mass hierarchy, if parameters are favorable. T2K will benefit from the extended running time and the upgrade of the  near detectors. The next generation long baseline experiments JUNO (scintillator detector), T2HK (Hyper-Kamiokande water Cherenkov detector), and DUNE (liquid argon TPCs)  will have real chances to discover the mass hierarchy and CP violation with the sensitivity of $\geq 5\sigma$.  
 
\acknowledgments

This work was supported  by the RFBR/JSPS  grant \# 17-52-50038.

\end{document}